\documentclass[prb,aps,superscriptaddress,amsmath,amssymb,showpacs,twocolumn,floatfix]{revtex4}
\usepackage[pdftitle={Charge density wave surface phase slips and non-contact nanofriction}]{hyperref}
\usepackage{epsfig}
\usepackage{amsfonts}
\usepackage{amsmath}
\newcommand{\md}[1]{\left|#1\right|}
\newcommand{\FE}{\mathcal{F}}
\newcommand{\qq}{\mathbf{Q}}
\newcommand{\rr}{\mathbf{r}}
\begin{document}

\title{Theory of charge density wave non-contact friction}

\author{Franco Pellegrini}
\affiliation{SISSA, Via Bonomea 265, I-34136 Trieste, Italy}
\affiliation{CNR-IOM Democritos National Simulation Center, Via Bonomea 265, I-34136 Trieste, Italy}

\date{\today}

\begin{abstract}
A mechanism is proposed to describe the occurrence of distance-dependent dissipation peaks in the dynamics 
of an atomic force microscope tip oscillating over a surface characterized by a charge density wave state. 
The dissipation has its origin in the hysteretic behavior of the tip oscillations occurring at positions 
compatible with a localized phase slip of the charge density wave.
This model is supported through static and dynamic numerical simulations of the tip surface interaction
and is in good qualitative agreement with recently performed experiments on a NbSe$_2$ sample.
\end{abstract}


\maketitle

%%%%%%%%%%%%%%%%%%%%%%%%%%%%%%%%%%%%%%%%%%%%%%%%%%%%%%%
The study of the microscopic mechanisms leading to energy dissipation and friction has very important theoretical
and practical implications. In recent years, experiments have started to single out the effects of microscopic 
probes in contact or near contact with different surfaces, and much theoretical effort has been devoted 
to the full understanding of such experiments~\cite{Vanossi13}.
In particular, the minimally invasive non-contact experiments offer a chance to investigate delicate surface 
properties and promise to bring new insight on localized effects and their interaction with the bulk.

Recently, a non-contact atomic force microscopy (AFM) experiment~\cite{Kisiel13} on a NbSe$_2$ sample has shown  
dissipation peaks appearing at specific heights from the surface and extending up to $2$ nm far from it. These peaks
were obtained with tips oscillating both parallel and perpendicular to the surface, and in a range of temperatures
compatible with the surface charge density wave (CDW) phase of the sample.
In this paper, a model is proposed explaining in detail the mechanism responsible for these peaks: the tip 
oscillations induce a charge perturbation in the surface right under the tip, but, due to the nature of the CDW order
parameter, multiple stable charge configurations exist characterized by different ``topological'' properties.
When the tip oscillates at distances corresponding to the crossover of this different manifolds, the system is 
not allowed to follow the energy minimum configuration, even at the low experimental frequencies of oscillation,
and this gives rise to a hysteresis loop for the tip, leading to an increase in the dissipation.

While the idea behind this dissipation mechanism has been proposed by the author and collaborators in the original
paper~\cite{Kisiel13}, this article expands on the technical aspects of the model, highlighting details to appear
in a future publication.

%%%%%%%%%%%%%%%%%%%%%%%%%%%%%%%%%%%%%%%%%%%%%%%%%%%%%%%%%%%%
\section{The Model} In this article, the term CDW is used to indicate a periodic modulation of the charge density $\rho$,
irrespective of the process behind its generation. This modulation is described, in the unperturbed system and
for the simplest form of CDW, 
as $\rho(\rr)=\rho_0\cos(\qq\cdot\rr)$, where $\rho_0$ is the intensity and $2\pi Q^{-1}$ the characteristic wavelength.
Perturbations to CDWs have been studied extensively~\cite{Gruner88,FukLee78,LeeRice79,Tucker89}, but most studies 
are concerned either with uniform perturbations (e.g. an external electric field) or point-like perturbations 
(e.g. pinning by defects), and often consider one-dimensional models, appropriate for quasi-one-dimensional materials,
where the coherence length in the perpendicular directions is smaller than the atomic distance. 
What is considered here, instead, is the effect of a localized but extended perturbation of typical length scale similar to 
the CDW wavelength, acting on a material where the coherence length is macroscopic in more than one dimension.

Starting from the standard Fukuyama-Lee-Rice model~\cite{FukLee78,LeeRice79} for CDW, the charge modulation is 
described through a Ginzburg-Landau theory as a classical elastic medium. 
The complex order parameter $\psi(\rr)=A(\rr)e^{i\phi(\rr)}$ is considered, taking into account the amplitude degree 
of freedom $A$, as well as the phase $\phi$. The charge density can be expressed as $\rho(\rr)=\rho_0A(\rr)\cos(
\qq\cdot\rr+\phi(\rr))$, so for the unperturbed system $A(\rr)=1$ and $\phi(\rr)=\phi_0$ are constant (at 
least locally), and the free energy reads:
\begin{equation} \label{eq:FE0}
\FE_0[\psi(\rr)]=\int\left[-2f_0\md{\psi(\rr)}^2+f_0\md{\psi(\rr)}^4+\kappa\md{\nabla\psi(\rr)}^2\right]\mathrm{d}\rr,
\end{equation}
where $f_0$ sets the energy scale and $\kappa$ represents the elastic energy contribution.

Considering the external perturbation by the AFM tip, generally described as a potential $V(\rr)$ 
coupling to the charge density, adds an interaction term to the free energy:
\begin{equation} \label{eq:FEV}
\FE_V[\psi(\rr)]=\int\left[V(\rr)\mathrm{Re}(\psi(\rr)e^{i\qq\cdot\rr})\right]\mathrm{d}\rr.
\end{equation}
A well studied case is the impurity one~\cite{Tucker89,TutZaw85}, where $V(\rr)=\sum_i\delta(\rr-\rr_i)$. In that case
phase oscillations are often enough to describe the ground state of the system, which comes from the balance of 
elastic and potential energy. 
These point-like perturbations, however, only impose a likewise point-like constraint on the order parameter, and
therefore cannot lead to a phase slip in the absence of an external driver~\cite{Tucker89}.
I will therefore consider the case where $V(\rr)$ has a finite width of the order of the wavelength $2\pi Q^{-1}$,
and minimize the total free energy $\FE=\FE_0+\FE_V$ given a specific shape of $V(\rr)$. 
Following in the steps of the impurity model and considering only phase perturbations, the functional would be
\begin{equation} \label{eq:FEphi}
\FE[\phi(\rr)]=\int\left[\kappa\md{\nabla\phi(\rr)}^2+V(\rr)\rho_0\cos(\qq\cdot\rr+\phi(\rr))\right]\mathrm{d}\rr.
\end{equation}
This model, however, is inadequate to describe this specific effect. In fact, considering a purely one-dimensional 
model would lead to a linear behavior of $\phi$ where the potential is zero. Since we expect a decay far from the 
perturbation, this is a clearly unphysical result. 
Moreover, due to the nature of the phase, which is defined modulo $2\pi$, given some boundary conditions the solution
is not univocally defined unless the total variation of $\phi$ is also specified. Assuming the phase to have the
unperturbed value $\phi_0$ far from the perturbation, we can define the integer \textit{winding number} $N$ of 
a solution as the integral
\begin{equation} \label{eq:WN}
N=\frac{1}{2\pi}\int\nabla\phi(x)\mathrm{d}x,
\end{equation}
taken along the CDW direction $\qq$ (with $N=0$ typically representing the unperturbed case).
Since any change in the winding number along the $\qq$ direction would extend to the whole sample and unnaturally 
raise the energy of such a solution, to recover a physical result one needs to take into account the amplitude 
degree of freedom, which will allow for the presence of dislocations and local changes in the winding number.

For these reasons, the minimization will be performed in two dimensions, with the complete complex order parameter
and in a subspace with a defined winding number.
The final result is expected to be similar to what previously considered in the wider context of phase 
slip~\cite{Tucker89} and more specifically in the case of localized phase slip centers~\cite{Maki95,Gorkov84}. 
Namely, the local strain induced by the perturbation on the phase will reduce the order parameter amplitude,
to the point where a local phase slip event becomes possible. In more than one dimension, the boundary 
between areas with different winding number will be marked by structures such as \textit{vortices}.

From this preliminary analysis, the mechanism responsible for the dissipation peaks can be understood: as the 
tip approaches the surface, it encounters points where the energies of solutions with different winding number 
undergo a crossover. At these points the transition between manifolds is not straightforward, due to the mechanism 
required to create the vortices, therefore the oscillations lead to jumps between different manifolds, resulting 
in hysteresis for the tip and ultimately dissipation.

%%%%%%%%%%%%%%%%%%%%%%%%%%%%%%%%%%%%%%%%%%%%%%

\section{Simulations} 
To asses the validity of the proposed mechanism, numerical simulations of the tip-surface interaction were performed.
A two-dimensional model is considered, since it takes into account the relevant effects while keeping the simulation 
simpler; moreover, the experimental substrate has a quasi-two-dimensional structure, so that volume effects can be
expected to be negligible. Differently from the experimental system~\cite{McMill75}, the CDW is characterized by a 
single wave vector $\qq$, leading to a simpler order parameter and a clearer effect.
To represent the effect of the tip, the shape of a van der Waals potential $C/r^6$ is integrated over a conical 
tip at distance $d$ from the surface. The result can be reasonably approximated in the main area under the tip by 
a Lorentzian curve
\begin{equation} \label{eq:ExtPot}
V(\rr;d)=\frac{V_0(d)}{\rr^2+\sigma(d)^2} ,
\end{equation}
where $\rr$ is the distance in the plane from the point right below the tip and the parameters are found to scale like 
$V_0(d)=\bar{V}/d$ and $\sigma(d)=\bar{\sigma}d^2$.

%%%%%%%%%%%%%%%%%%%%%%%%%%%%%%%%%%%%%%%%%%%%%%%%%%%%%%%%%%%%%%%%%%%%%%%%%%%%%%
\begin{figure}[!tb]\centering
\includegraphics[width=.48\textwidth]{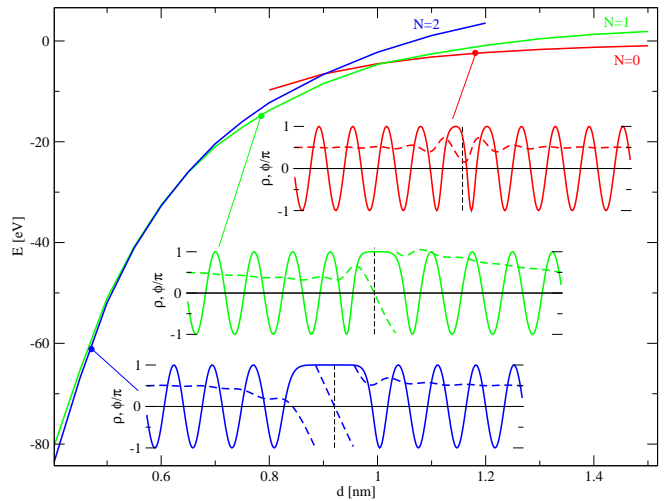}
\caption{\label{fig:encurves}Free energy $\FE$ as a function of tip distance $d$ for subspaces with different 
winding number $N$.
Results from simulations on a 201$\times$201 grid with parameters (see text) $f_0=10$~eV/nm, $\kappa=1$~eV, 
$\qq=2.5$~nm$^{-1}$, $\bar{V}=−9.4$~eV$\cdot$nm, $\sigma=1.2$~nm$^{-1}$ and boundary conditions $\psi_0=i$.
Insets: charge density $\rho$ (full lines) and phase $\phi$ (dashed lines) along the line passing right below the tip
(indicated by the vertical dashed line) for different winding number $N$, at the positions indicated by the dots
on the energy curves.} 
\end{figure}
%%%%%%%%%%%%%%%%%%%%%%%%%%%%%%%%%%%%%%%%%%%%%%%%%%%%%%%%%%%%%%%%%%%%%%%%%%%%%%%

Knowing the shape of the perturbation, the total free energy $\FE=\FE_0+\FE_V$ is minimized numerically
on a square grid of points with spacing much smaller than the characteristic wavelength of the CDW, 
imposing a constant boundary condition $\psi_0$ on the sides perpendicular to $\qq$, while setting 
periodic boundary conditions in the other direction to allow for possible phase jumps. 
The minimization is carried out through a standard conjugated gradients algorithm~\cite{NumRec}.
The parameters employed are order of magnitude estimates of the real parameters, which reproduce the relevant 
experimental effects in a qualitative fashion.

The charge density $\rho$ and phase $\phi$ profile along the line passing right below the tip is shown in the 
insets of Fig.~\ref{fig:encurves} for solutions with different winding number $N$. The position of the 
attractive tip is indicated by the vertical dashed line, and the phase slip occurred under it, leading to
and increase in charge density, is clearly visible. Parallel lines far from the tip always revert to the 
$N=0$ manifold, through the creation of vortices.

The main curves in Fig.~\ref{fig:encurves} represent the minimum energy at defined $N$ as a function of the distance $d$. 
Since the solution with a given winding number lies in a local minimum, it is possible to use the minimization algorithm
to find solutions in a certain subspace, even when this is not the global minimum for a specific distance, by starting 
from a reasonable configuration (namely, the minimum at close distance).
Two different crossing points can be seen, which would give rise to two peaks in the experimental dissipation
trace. Of course a more complex CDW configuration or different parameters would give rise to more peaks.

%%%%%%%%%%%%%%%%%%%%%%%%%%%%%%%%%%%%%%%%%%%%%%%%%%%%%%%%%%%%%%%%%%%%%%%%%%%%%%
\begin{figure}[!b]\centering
\includegraphics[width=.48\textwidth]{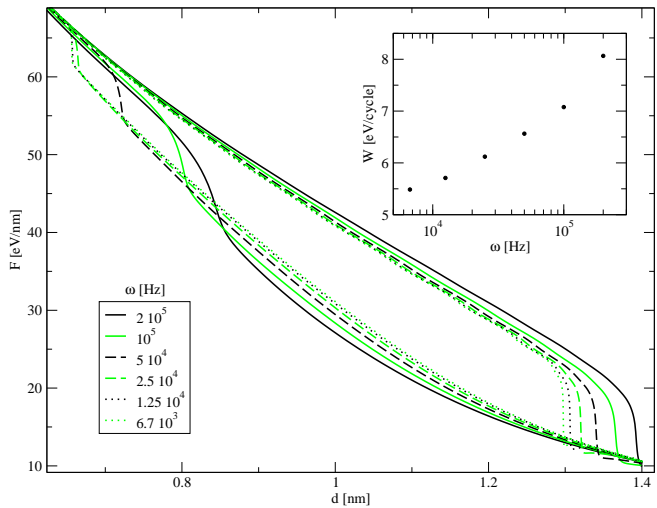}
\caption{\label{fig:fcurves}Force as a function of distance for evolutions with $d_0=1$~nm, $\bar{d}=0.4$~nm and different 
values of $\omega$ with $\Gamma=10^{-7}$~eV$\cdot$s. Inset: total work $W$ as a function of oscillation frequency $\omega$.}
\end{figure}
%%%%%%%%%%%%%%%%%%%%%%%%%%%%%%%%%%%%%%%%%%%%%%%%%%%%%%%%%%%%%%%%%%%%%%%%%%%%%%%

To completely justify the validity of the dissipation mechanism, we need to look into the dynamics of the CDW, to guarantee
that the evolution through a crossing point does not lead to immediate relaxation between different manifolds. To do this,
the time evolution of the system was simulated, following the time-dependent Ginzburg-Landau equation~\cite{Gorkov84}
\begin{equation} \label{eq:TDGL}
-\Gamma\frac{\partial\psi}{\partial t}=\frac{\delta\FE}{\delta\psi^*}.
\end{equation}
This equation can be interpreted as an overdamped relaxation, with a coefficient $\Gamma$, of the order parameter towards 
the equilibrium position. Integrating this equation (through a standard Runge-Kutta algorithm~\cite{NumRec}), the force as
a function of the distance can be computed for a tip performing a full oscillation perpendicular to the surface according 
to the law $d(t)=d_0+\bar{d}\cos(\omega t)$.
Fig.~\ref{fig:fcurves} shows the force evolution during such oscillations at different frequencies. As we can see the tip
suffers a hysteresis even at low frequencies, since the decay from one manifold to the other happens far from the crossing
point. The area of the loops represents directly the dissipated energy per cycle $W$, as reported in the inset.

%%%%%%%%%%%%%%%%%%%%%%%%%%%%%%%%%%%%%%%%%%%%%%%%%%%%%%%%%%%%%%%%%%%%%
\section{Conclusion} Based on these simulations, the theory hereby presented is shown to be compatible with the experimental 
findings: it accounts for the existence of multiple peaks, their appearance far away from the surface and their nature 
being related to the CDW structure of the sample.

It is interesting to notice that the vortices appearing in our simulations have been described before in the context of CDW 
conduction noise \cite{OngVerMaki84}, where their creation and movement justifies the phase slip near the CDW boundaries. 
In this sense, this theory lies in between these macroscopic effect and the simple one-dimensional model of defect pinning 
and phase slip \cite{Tucker89}, as is appropriate for a localized but extended perturbation.

To conclude, I have presented a mechanism to explain peaks in the dissipation of a tip oscillating
at specific distances above a CDW surface: these occur around instability points corresponding to the crossing of energy levels 
characterized by different winding numbers. Numerical simulations in a system of reduced complexity support the validity of 
this mechanism. 

It would be interesting to investigate the same effect in other systems displaying CDW or even spin density waves, as well as
systems where the origin of charge modulation is related to the Fermi surface and not to other effects.

%%%%%%%%%%%%%%%%%%%%%%%%%%%%%%%%%%%%%%%%%%%%%%%%%%%%%%%%%%%%%%%%
\textit{Acknowledgements} --- The author thanks his collaborators G.E. Santoro and E. Tosatti. He acknowledges research support by SNSF, 
through SINERGIA Project CRSII2 136287/1, by ERC Advanced Research Grant N. 320796 MODPHYSFRICT, and by MIUR, through 
PRIN-2010LLKJBX\_001
%%%%%%%%%%%%%%%%%%%%%%%%%%%%%%%%%%%%%%%%%%%%%%%%%%%%%%%%%%%%%%

\end{document}